\documentclass[showpacs,preprint,superscriptaddress]{revtex4-1}
\usepackage{amsmath}    
\usepackage{amssymb}
\usepackage{wasysym}
\usepackage{mathtools}
\usepackage{graphicx}   
\usepackage{verbatim}   
\usepackage{color}      
\usepackage{subfigure}  
\usepackage{hyperref}   
\usepackage{float}

\begin{document}

\title{Lovelock black holes surrounded by quintessence}

\author{Sushant G. Ghosh}
\email{sgghosh@gmail.com,sghosh2@jmi.ac.in}
\affiliation{Astrophysics and Cosmology Research Unit, School of Mathematics, Statistics and Computer Science, University of KwaZulu-Natal, Private Bag X54001, Durban 4000, South Africa}
\affiliation{Centre for Theoretical Physics, and  Multidisciplinary Centre for Advanced Research and Studies (MCARS), Jamia Millia Islamia, New Delhi 110025, India}

\author{Sunil D. Maharaj}
\email{maharaj@ukzn.ac.za}
\affiliation{Astrophysics and Cosmology Research Unit, School of Mathematics, Statistics and Computer Science, University of KwaZulu-Natal, Private Bag X54001, Durban 4000, South Africa}

\author{Dharmanand Baboolal}
\email{Baboolald@ukzn.ac.za}
\affiliation{Astrophysics and Cosmology Research Unit, School of Mathematics, Statistics and Computer Science, University of KwaZulu-Natal, Private Bag X54001, Durban 4000, South Africa}

\author{Tae-Hun Lee}
\email{taehunee@gmail.com}
\affiliation{Astrophysics and Cosmology Research Unit, School of Mathematics, Statistics and Computer Science, University of KwaZulu-Natal, Private Bag X54001, Durban 4000, South Africa}
\date{\today}
\begin{abstract}
Lovelock gravity consisting of the dimensionally continued Euler densities is a natural generalization of general relativity to higher dimensions such that equations of motion are still second order, and the theory is free of ghosts. A scalar field with a positive potential that yields an accelerating universe has been termed quintessence.  We present exact black hole solutions in $D$-dimensional Lovelock gravity surrounded by quintessence matter and also perform a detailed thermodynamical study. Further, we find that the mass, entropy and temperature of the black hole are corrected due to the quintessence background. In particular, we find that phase transition occurs  with divergence of heat capacity at the critical horizon radius, and that specific heat becomes positive for $r_h<r_c$ allowing the black hole to become thermodynamically stable. 
\end{abstract}

\maketitle
\section{Introduction}
High-precision observational data have confirmed  evidence that the Universe is undergoing a phase of accelerated expansion \cite{rp} which may be due to dark energy that is gravitationally self-repulsive. Quintessence, a time-evolving, spatially inhomogeneous component with negative pressure  is a possible candidate \cite{ratra} for dark energy, which is characterized by the equation of state $P=w\rho$, where $P$ is the pressure, $\rho$ is the energy density, and  $-1 < w < -1/3$ .  It is an exotic kind of field present everywhere in the universe that exerts force so that particles move away from each other by overpowering gravity and the other fundamental forces. The fact that dark energy constitutes 70\% of the universe and black holes are part of our universe, makes the study of black holes surrounded by dark energy important.  A spherical symmetry solution to  the Einstein equations surrounded by quintessence matter was first obtained by Kiselev \cite{Kiselev:2002dx} which includes the black hole charged or not in flat or de Sitter space.
  
Thereafter, significant attention has been devoted to discussion of  static spherically symmetric black hole solutions surrounded by quintessence matter and their properties \cite{Kiselev:2002dx,Ma:2007zze,Fernando:2012ue,Feng:2014yjj,Malakolkalami:2015cza,Hussain:2014fca}. Also, several extensions of the Kiselev solutions \cite{Kiselev:2002dx} have been obtained. These include models for a charged black hole \cite{Azreg-Ainou:2014lua}, the Nariai  solution \cite{FERNANDO:2013uxa,Fernando:2014wma} and extensions to higher dimensions \cite{Chen:2008ra}.  These black hole solutions are possible only under special choice of  parameter in the energy momentum tensor of quintessence, depending on the state parameter $w$. The black hole thermodynamics for the quintessence corrected black hole solutions were discussed in \cite{Wei:2011za,Thomas:2012zzc,AzregAinou:2012hy,Tharanath:2013jt,Ghaderi:2016dpi} and the quasinormal modes of such solutions have been obtained \cite{Zhang:2007nu,Varghese:2008ky,Saleh:2011zz,Tharanath:2014uaa}.  Owing to the quintessence surrounding the  black hole, the thermodynamic quantities have also been corrected except for the black hole entropy, and it is shown that phase transition is achievable. Ghosh \cite{Ghosh:2015ovj} and Toshmatov {\it et al.} \cite{Toshmatov:2015npp} further generalized the Kiselev solution to include  the axially symmetric case, i.e.,  the Kerr-like black hole was also addressed.   It was shown that  a rotating counterpart of the Kiselev solutions \cite{Kiselev:2002dx}  can be  identified for quintessence parameter $ w=1/3 $, exactly as the Kerr\(-\)Newman black hole and as the Kerr black hole according to a choice of the integration constant \cite{Ghosh:2015ovj,Toshmatov:2015npp}.

The accelerated expansion of the Universe has also inspired  several modifications of general relativity which aim to explain the cosmic acceleration and reconstruct the entire expansion history. A natural modification of general relativity is the Lovelock gravity, whose Lagrangian consists of the dimensionally extended Euler densities. This was obtained by Lovelock in an attempt to obtain the most general tensor that satisfies properties of the Einstein tensor in higher dimensions \cite{Lovelock:1971yv}. The Lovelock action contains higher order curvature terms and  reduces to the Einstein-Hilbert action in four-dimensions, and its second order term is the Gauss-Bonnet invariant. 
The Lovelock theories have some special characteristics, amongst the larger class of general higher-curvature theories, in having field equations involving not more than second derivatives of the metric. They have also been shown to be free of ghosts when expanding about flat space, evading any problems with unitarity. Exact solutions describing black holes have also been found for these theories  \cite{Boulware:1985wk,jtw,Wiltshire}, and  later several generalizations of the Boulware-Desser solution  have also been  discussed \cite{som1,hr,sgr,sgsm,Ghosh:2014pga,Ghosh}.  
Recently, the quintessence atmosphere to the Einstein-Gauss-Bonnet black hole has been analyzed by Ghosh {\it et al.} \cite{Ghosh:2016ddh} . In this analysis the case of no quintessence $\omega=0$ reduces to the Boulware and Deser \cite{Boulware:1985wk,jtw} Gauss-Bonnet black hole solution, and for quintessence parameter $\omega=1/2$ with an appropriate choice of the integration constant it is mathematically similar to the the charged Gauss-Bonnet black hole case due to Wiltshire \cite{Wiltshire}. 
The spherically symmetric vacuum solutions to Lovelock gravity have been found independently in \cite{Cai:2003kt, Myers:1988ze}, and also recent work includes a class of Lovelock black holes with conformal scalar hair \cite{Hennigar:2016ekz,Hennigar:2016xwd,Giribet:2014bva}. It is the purpose of this paper to model the possible effect of surrounding  quintessence matter on the spherically symmetric black hole solution in Lovelock gravity. In particular, we can explicitly explain how the effect of a background quintessence matter can alter black hole solutions and their thermodynamics. 

The paper is organized as follows. In Sec. II, we begin by reviewing the higher dimensional  \cite{Chen:2008ra}  spherically symmetric black hole surrounded by quintessence matter.  We have analyzed the quintessence background  of  Kiselev \cite{Kiselev:2002dx} to derive the resulting exact Lovelock black holes. The thermodynamic properties of the new derived solutions are explored and we obtain the thermodynamic quantities exactly in Sec III .  In Sec. IV a discussion covering   the thermodynamical stability of black holes is presented, and Sec.~V gives the concluding remarks.

We use units which fix the speed of light and the gravitational constant via $8\pi G = c = 1$, and use 
the metric signature ($-,\;+,\;+,\; \cdots,\;+$).

\section{D-dimensional black holes surrounded by quintessence.}
In this section, we review  $D$-dimensional static spherically symmetric black holes surrounded by quintessence matter. The metric for general static spherically symmetric spacetime in $D$-dimensions can be written as \cite{Chen:2008ra}
\begin{equation}
ds^2=-f(r)dt^2+\frac{1}{f(r)} dr^2+r^2 d\Omega^2_{D-2},\label{dsf}
\end{equation}
where $d\Omega_{D-2}$ is a line element on a $D-2$ dimensional hypersurface with constant scalar curvature $\kappa=1,0$ and $-1$, respectively for spherical, flat and hyperbolic spaces. For $D$-dimensional spherically symmetric spacetime, the energy momentum tensor can be written as  \cite{Chen:2008ra}
\begin{equation}
T_t{}^t=\chi(r),~~T_t{}^b=0,~~T_a{}^b=\xi(r)r_ar^b+\eta(r)\delta_a{}^b.\label{Symmetric Energy Momentum}
\end{equation}
Isotropic average over the angles in Eq.~(\ref{Symmetric Energy Momentum}) leads to \cite{Kiselev:2002dx,Chen:2008ra}
\begin{equation}
\langle T_a{}^b\rangle=\zeta(r)\delta_a{}^b,~~\zeta(r)=-\frac{1}{D-1}\xi(r)r^2+\eta(r),
\end{equation}
and for quintessence matter we have 
\begin{equation}
\zeta(r)=-\omega_q\chi(r).
\end{equation}
Then the general expression of the energy momentum tensor for quintessence in $D$-dimensional spacetime \cite{Chen:2008ra} is
\begin{equation}
T_t{}^t=\rho_q(r),~~T_a{}^b=\rho_q(r)\alpha\left\{-\big[1+(D-1)\eta\big]\frac{r_ar^b}{r_n r^n}+\eta\delta_a{}^b\right\}.\label{Tmn}
\end{equation}
Again, if one takes isotropic average over the angles,
\begin{equation}
\langle r_a r^b\rangle=\frac{r_n r^n\delta_a{}^b}{(D-1)},\label{rAverage}
\end{equation}
one obtains
\begin{equation}
\langle T_a{}^b\rangle=\rho_q \frac{\alpha}{D-1}\delta_a{}^b=-p_q\delta_a{}^b.\label{TAverage}
\end{equation}
Using the equation of state for quintessence matter \cite{Kiselev:2002dx,Chen:2008ra}
\begin{equation}
p_q=\omega_q\rho_q,
\end{equation}
we obtain
\begin{equation}
\omega_q=\frac{\alpha}{(D-1)}.
\end{equation}
Clearly for quintessence $-1<\omega_q<0$ leads to $-(D-1)<\alpha<0$.

The Einstein field equations, $G_\nu{}^\mu=T_\nu{}^\mu$ lead to the following equations:
\begin{eqnarray}\label{Tttf}
T_t{}^t&=&T_r{}^r=-\frac{D-2}{4r^2}\big[f^\prime+(D-3)(f-1)\big],\\
T_{\theta_1}{}^{\theta_1}&=&8\pi T_{\theta_2}{}^{\theta_2}=\cdots 8\pi T_{\theta_{D-2}}{}^{\theta_{D-2}}\\\nonumber\label{Tthetaf}
&=&-\frac{1}{4r^2}\big[r^2f^{\prime\prime}+2(D-3)rf^\prime+(D-4)(D-3)(f-1)\big],
\end{eqnarray}
where we have chosen units $8\pi G=1$ and assumed that for the black hole $T_t{}^t=T_r{}^r$.
From Eqs.~(\ref{Tmn}) and (\ref{Tttf}), one obtains
\begin{equation}
\eta=-\frac{(D-1)\omega_q+1}{(D-1)(D-2)\omega_q},
\end{equation}
and the energy momentum tensors takes the form
\begin{eqnarray}\label{Trho}
T_t{}^t=T_r{}^r&=&\rho_q,\\
T_{\theta_1}{}^{\theta_1}=T_{\theta_2}{}^{\theta_2}=\cdots T_{\theta_{D-2}}{}^{\theta_{D-2}}&=&-\frac{1}{(D-2)}\rho_q\big[(D-1)\omega_q+1\big].\label{Tthetarho}
\end{eqnarray}
From Eqs.~(\ref{Tttf}), (\ref{Tthetaf}), (\ref{Trho}) and (\ref{Tthetarho}), we get a master differential equation for $f:$
\begin{equation}
r^2f^{\prime\prime}+[(D-1)\omega_q+2D-5]rf^\prime+(D-3)[(D-1)\omega_q+D-3](f-1)=0.\label{eqf}
\end{equation}
Eq.~(\ref{eqf}) admits the solution
\begin{equation}
f(r)=1-\frac{\mu}{r^{D-3}}+\frac{b}{r^{(D-1)\omega_q+D-3}}, \label{fsol}
\end{equation}
where $\mu$ and $b$ are normalized factors.
From Eqs.~(\ref{Tttf}), (\ref{Trho}) and (\ref{fsol}) the energy density for quintessence is
\begin{equation}
\rho_q=-\frac{\omega_qq(D-1)(D-2)}{2 r^{(D-1)(\omega_q+1)}},\label{rhoq}
\end{equation}
where we have chosen $b$ as $-q$ to be negative for positive energy density. The metric for a spherically symmetric black hole surrounded by quintessence reads \cite{Chen:2008ra}
\begin{equation}
ds^2=-\left[1-\frac{\mu}{r^{D-3}}-\frac{q}{r^{(D-1)\omega_q+D-3}}\right]dt^2+\left[1-\frac{\mu}{r^{D-3}}-\frac{q}{r^{(D-1)\omega_q+D-3}}\right]^{-1}dr^2+r^2d\Omega_{D-2}. \label{ds}
\end{equation}
Note that this metric depends not only on the dimension $D$ but also on a quintessence state parameter $\omega_q$. In the limit $q\rightarrow 0$, the metric goes to the Schwarzschild-Tangherlini metric in $D$ dimensions \cite{Tangherlini:1963bw}
\begin{equation}
ds^2=-\left[1-\frac{\mu}{r^{D-3}}\right]dt^2+\left[1-\frac{\mu}{r^{D-3}}\right]^{-1}dr^2+r^2d\Omega_{D-2}.
\end{equation}
When $\omega_q=(D-3)/(D-1)$, the metric in Eq.~(\ref{ds}) reduces to the $D$-dimensional Reissner-Nordstr\"{o}m metric with $q$ replaced with $-Q^2$ \cite{Reissner:1916, Nordström:1918}:
\begin{equation}
ds^2=-\left[1-\frac{\mu}{r^{D-3}}+\frac{Q^2}{r^{2(D-3)}}\right]dt^2+\left[1-\frac{\mu}{r^{D-3}}+\frac{Q^2}{r^{2(D-3)}}\right]^{-1}dr^2+r^2d\Omega_{D-2}.
\end{equation} 
The metric for a black hole in the string cloud given in \cite{Ghosh:2014pga, Lee:2014dha} can be obtained from Eq.~(\ref{ds}) with $q=-a$ and $\omega_q=-1/(D-1)$
as 
\begin{equation}
ds^2=-\left[1-\frac{\mu}{r^{D-3}}+\frac{a}{r^{D-4}}\right]dt^2+\left[1-\frac{\mu}{r^{D-3}}+\frac{a}{r^{D-4}}\right]^{-1}dr^2+r^2d\Omega_{D-2},
\end{equation}
where $a$ is a constant from energy momentum tensor of a string cloud.
\section{Lovelock back holes surrounded by quintessence matter}
Among the higher derivative gravities the Lovelock gravity is very special \cite{Myers:1988ze}, as the field equations are still second order, and this theory appears as the low energy limit of string theories \cite{Gross:1986iv}. We want to model black holes surrounded by quintessence matter in the general Lovelock gravity. The Lovelock gravity consists of Euler densities \cite{Lovelock:1971yv, Myers:1988ze, Ghosh:2014pga}
\begin{equation}
S=S_L+S_M,\label{action}
\end{equation}
where $S_M$ is the action for Quintessence matter which is as an alternative to a positive cosmological constant, can realize a fluid with a equation of state $\omega > -1$. A general action for quintessence in D-dimensional spacetime is  
\begin{equation}\label{Qaction}
\int d^D x \sqrt{-g} \left[ - \frac{1}{2} (\nabla \phi)^2 - V(\phi)\right]. 
\end{equation}
The theory of the quintessence has been put forward as an alternative to a positive cosmological constant. The simplest way to generate models of quintessence is to adopt a matter system described by a single, real scalar field entering the action with a canonical kinetic
term and an exponential potential of the form  $V(\phi) = c \exp(\gamma \phi)$, with $c,\gamma >0$. A simple calculation leads to 
\begin{equation}\label{omega}
\omega_q = -1 + \frac{(D-2)\gamma^2}{(D-1)2}
\end{equation}
when we choose $\gamma^2=2$, one obtains
\begin{equation}
\omega_q = - \frac{1}{(D-1)}
\end{equation}
and when $D=4$, one gets $\omega_q = - 1/3$. The  $S_L$ in $D$ spacetime dimension reads 
\begin{equation}
S_L=\frac{1}{2}\int dx^{D}\sqrt{-g}\mathcal{L},
\end{equation}
with Lagrangian $\mathcal{L}$ 
\begin{equation}
\mathcal{L}=\alpha_0+R+\sum\limits_{s=2}^{[(D-1)/2]}\alpha _{s}\mathcal{L}_s,
\end{equation}
where $\alpha_s$ is a coupling constant, $\mathcal{L}_s$ is the Euler density given by
\begin{equation}
\mathcal{L}_s=\frac{1}{2^{s}}\delta _{\alpha _{1}\beta_{1}...
\alpha _{s}\beta _{s}}^{\mu _{1}\nu _{1}...\mu _{s}\nu_{s}}
R^{\alpha _{1}\beta _{1}}{}_{\mu _{1}\nu _{1}}\cdots R^{\alpha _{s}\beta _{s}}{}_{\mu _{s}\nu _{s}}.\label{EulerDensity}
\end{equation}
The generalized Kronecker delta function is totally antisymmetric in all indices. The term $[(D-1)/2]$ on the summation means the integral part of $(D-1)/2$. Here $\mathcal{L}_2$ corresponds to the Gauss-Bonnet term \cite{Myers:1988ze}. Varying the action in Eq.~(\ref{action}) with respect to the metric tensor $g_{\mu\nu}$ we obtain the Lovelock field equations written in terms of generalized Einstein tensor \cite{Myers:1988ze,Cai:2003kt}
\begin{equation}
G_{\mu\nu}\equiv -\frac{\alpha_0}{2}g_{\mu\nu}+G_{\mu\nu}^{E}+\mathcal{G}_{\mu\nu}=T_{\mu\nu}, \label{Einstein eq}
\end{equation}
where 
\begin{equation}
G_{\mu\nu}=\frac{2}{\sqrt{-g}}\frac{\delta S_L}{\delta g^{\mu\nu}},~~T_{\mu\nu}=-\frac{2}{\sqrt{-g}}\frac{\delta S_M}{\delta g^{\mu\nu}}.
\end{equation}
The model is governed by the energy momentum tensor  $T_{\mu\nu}$ associated with quintessence matter defined in Eq.~(\ref{Trho}). $G_{\mu\nu}^{E}$ is the Einstein tensor, i.e.
\begin{equation}
G_{\mu\nu}^{E}=R_{\mu\nu}-\frac{1}{2}g_{\mu\nu}R.
\end{equation}
The Riemann-Lovelock curvature tensor $G_\mu{}^{\nu(s)}$ is given as \cite{Kastor:2006vw,Myers:1988ze,Ghosh:2014pga}
\begin{equation}
\mathcal{G}_\mu{}^{\nu}=\sum\limits^{m}_{s=2}\alpha_s\delta _{\mu\alpha _{1}\cdots\alpha _{s}\beta_{1}\cdots\beta _{s}}^{\nu\mu _{1}\cdots\mu_{s}\nu _{1}\cdots\nu_{s}}R^{\alpha _{1}\beta _{1}}{}_{\mu _{1}\nu _{1}}\cdots R^{\alpha _{s}\beta _{s}}{}_{\mu _{s}\nu _{s}}.
\end{equation}
It is observed that the field equations are of second order as the Lovelock gravity is the sum of dimensionally continued Euler densities. The field equations can be conveniently expressed as a polynomial equation \cite{Wheeler:1985qd}. The static spherically symmetric metric is given by Eq.~(\ref{dsf}) with
\begin{equation}
f(r)=\kappa-r^2F(r).\label{fF}
\end{equation}
Here $F(r)$ can determined by solving for the real roots of the following polynomial equation \cite{Myers:1988ze,Cai:2003kt}
\begin{equation}
P(F)=\sum\limits^{m}_{s=0}\tilde{\alpha}_s F^s=
\frac{2 M}{(D-2)\Sigma_{D-2}r^{D-1}}+\frac{q}{r^{(D-1)(\omega_q+1)}},\label{PF}
\end{equation}
where the coefficient $\tilde{\alpha}_s$ is defined by
\begin{eqnarray}\nonumber
\tilde{\alpha}_0&=&\frac{\alpha_0}{[(D-1)(D-2)]},~~~\tilde{\alpha}_1=1,\\
\tilde{\alpha}_s&=&\prod\limits_{i=3}^{2s}(D-i)\alpha_s,~~(s>1).\nonumber
\end{eqnarray}
Eqs.~(\ref{Trho}) and (\ref{rhoq}) have been used, $q$ is appropriately chosen such that $\omega_q\leq0$ and $M$ is an integration constant considered as the mass of a black hole, and $\Sigma_{D-2}$ is the volume of a $(D-2)$-unit sphere
\begin{equation}
\Sigma_{D-2}=\frac{2\pi^{(D-1)/2}}{\Gamma[(D-1)/2]}.
\end{equation}
Thus, Eq.~(\ref{dsf}), with Eqs.~(\ref{fF}) and (\ref{PF}), represents black holes in the Lovelock gravity surrounded by quintessence matter. 
The solution for $q=0$ was obtained by Cai \cite{Cai:2003kt} and the asymptotic behavior and causal structure of the solution have been analysed in detail \cite{Cai:2003kt}. Here we can also make a detailed analysis for the above black holes surrounded by quintessence matter on its horizon. When $q=0$ the  solution of $P(F)=0$ is a constant curvature vacua \cite{Myers:1988ze,Boulware:1985wk}. Again, for $q=0$, the cases $F_0<0$ and $F_0>0$ correspond to, respectively, anti-de Sitter and de Sitter background. In the limit $q\rightarrow0$ the solution becomes black hole vacuum solutions in the Lovelock gravity obtained in \cite{Myers:1988ze,Cai:2003kt}. 

\begin{figure}[H]
	\begin{center}
		\begin{tabular}{cc}
			\includegraphics[scale=1.0]{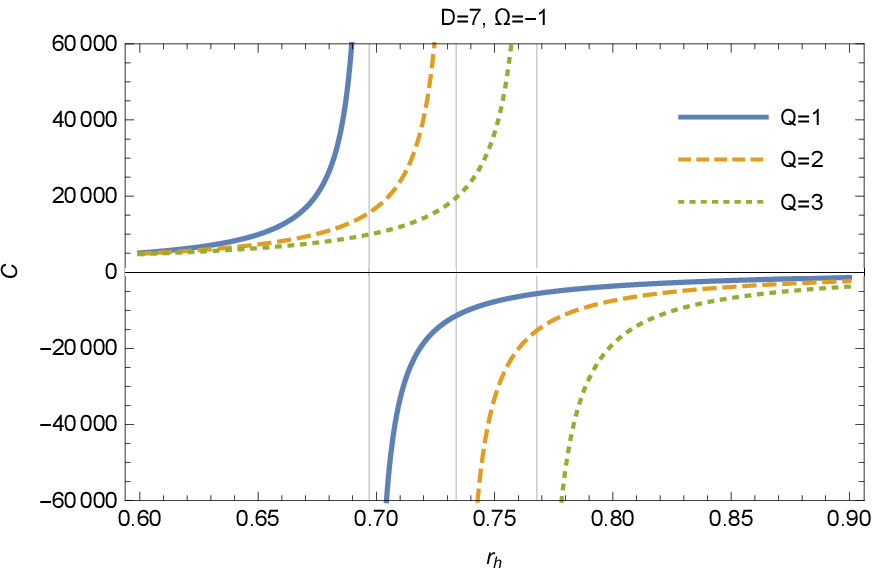} 
			& \includegraphics[scale=1.0]{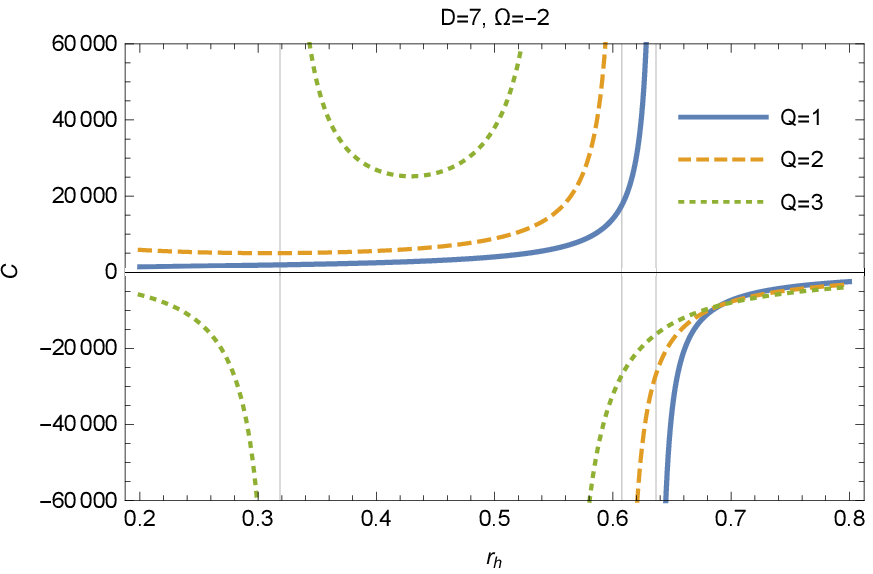}
		\end{tabular} 
	\end{center}
	\caption{The plots show how the heat capacity behaves for different values of $q$. We have chosen the parameters as $\kappa=1$, $D=7$, $\tilde{\alpha}_0=-1$ and $\tilde{\alpha}_1=\tilde{\alpha}_2=\tilde{\alpha}_3=1$. At $\Omega=-1$ the point $r_h=r_c$ where the heat capacity diverges decreses as $q$ increases while at $\Omega=-2$, $r_c$ increases as $q$ increases. For $\Omega=-2$ and $Q=3$ there exist two $r_c$s in $C$.}\label{figCq}
\end{figure}

\begin{figure}[H]
	\begin{center}
		\begin{tabular}{lll}
			\includegraphics[scale=1.0]{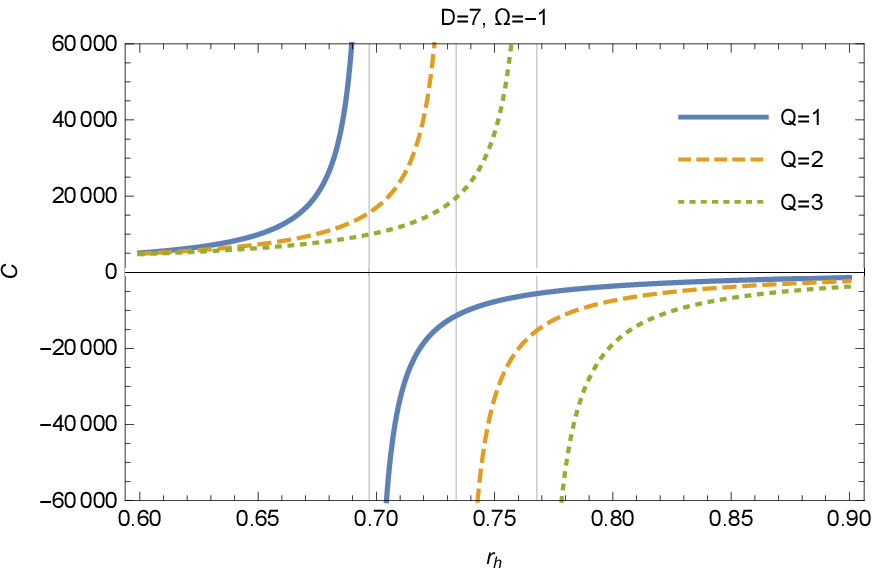} 
			&		\includegraphics[scale=1.0]{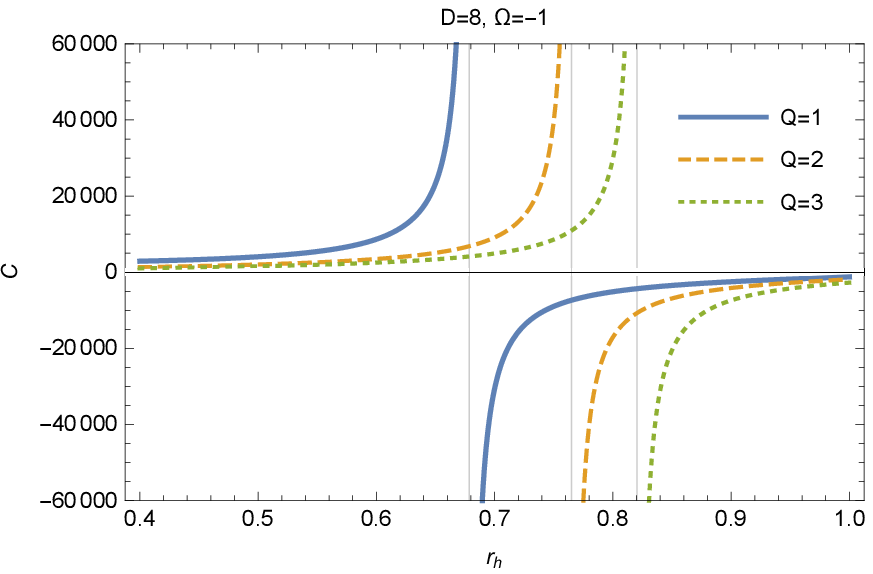}\\\includegraphics[scale=1.0]{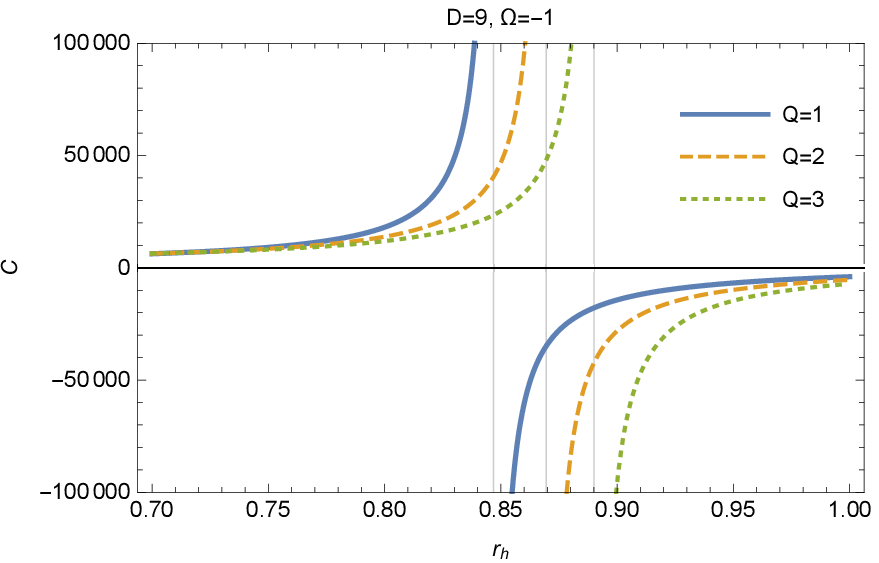}&\includegraphics[scale=1.0]{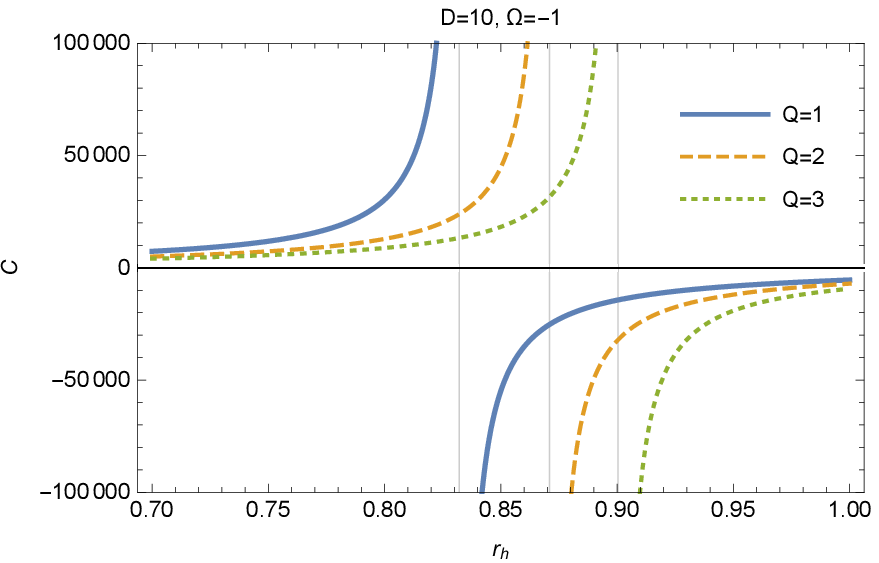}
		\end{tabular} 
	\end{center}
	\caption{The plots show how the heat capacity changes with different dimensions. We have chosen the parameters as $\kappa=1$, $\tilde{\alpha}_0=-1$ and $\tilde{\alpha}_1=\tilde{\alpha}_2=\tilde{\alpha}_3=1$.}\label{figD}
\end{figure}

Quintessence as one candidate for the dark energy is defined as an ordinary scalar field coupled to gravity. Black holes surrounded by dark energy are believed to play the crucial role in cosmology and one of the important characteristics of a black hole is its thermodynamical properties and also it  is interesting to know how does the dark energy affect the thermodynamics of the black holes in the Lovelock gravity. Having found the exact black hole solution, we can turn to discussion on black hole thermodynamics in analogy with \cite{Kastor:2006vw}, where we treat all dimension couplings as thermodynamic quantities. The black hole thermodynamics also provides insight into quantum properties of the gravitational field., in particular, the thermodynamics of AdS black holes has been of great interest since the pioneering work of Hawking and Page, which demonstrated the existence of a phase transition  in  AdS black holes cite{hp}.  The thermodynamical quantities associated with black holes can be expressed in terms of a horizon radius $r_h$ which satisfies $f(r_h)=0$ in Eq.~(\ref{fF}), leading to 
\begin{equation}
r_h^2=\frac{\kappa}{F(r_h)}.\label{rhF}
\end{equation}
The mass of the black hole in terms of horizons, by using the Eqs.~(\ref{PF}) and (\ref{rhF}), reads
\begin{equation}
M=\frac{(D-2)\Sigma_{D-2}}{2 }\left[\sum\limits^m_{s=0}\frac{\tilde{\alpha}_s\kappa^s}{r^{-(D-2s-1)}_h}-\frac{q}{r_h^{(D-1)\omega_q}}\right].\label{M}
\end{equation}
One can verify that in the limit $q\rightarrow 0$, it goes to the mass \cite{Myers:1988ze, Cai:2003kt}. It can be also seen that when $\tilde{\alpha}_s\rightarrow 0, (s\neq1)$ with $\kappa=1$, it goes to mass for the $D$ dimensional Schwarzschild-Tangherlini black hole \cite{Ghosh:2014pga}:
\begin{equation}
M=\frac{(D-2)\Sigma_{D-2}}{2 }r_h^{D-3}.
\end{equation}
Next we calculate thermodynamic quantities associated with black holes. The Hawking temperature associated with black holes is defined as $T=K/2\pi$, where $K$ is the surface gravity, which leads to 
\begin{equation}
T=\frac{f^\prime(r)}{4\pi},
\end{equation}
which on Eq.~(\ref{fF}), reads
\begin{equation}
T=\frac{1}{4\pi N(r_h)} \left[{\sum\limits^m_{s=0}\frac{\tilde{\alpha}_s\kappa^{s}(D-2s-1)}{r^{2s+2}_h}+\frac{q\omega_q(D-1)}{r_h^{(D-1)\omega_q+D+1}}}\right], \label{T}
\end{equation}
with $$N(r_h)={\sum\limits^m_{s=1}\frac{\tilde{\alpha}_s s\kappa^{s-1}}{r^{2s+1}_h}}.$$
It can be seen that in the limit $q\rightarrow 0$, it reduces to the Hawking temperature obtained in \cite{Cai:2003kt}. Whereas in the limit $\tilde{\alpha}_1\rightarrow1$ and $\tilde{\alpha}_s\rightarrow 0, (s\neq1)$ with $\kappa=1$ the temperature reduces to 
\begin{equation}
T=\frac{1}{4\pi}\left[\frac{(D-3)}{r_h}+\frac{q\omega_q(D-1)}{r_h^{(D-1)\omega_q+D-2}}\right].\label{TSchwarz}
\end{equation}
Eq.~(\ref{TSchwarz}) is the Hawking temperature in the $D$ dimensional Schwarzschild-Tangherlini black hole surrounded by quintessence matter \cite{Chen:2008ra,Ghosh:2014pga}. Further, when $q\rightarrow0$, it leads to the Hawking temperature for the $D$ dimensional Schwarzschild-Tangherlini black hole.  The Lovelock black holes obeys extended first law and Smarr
formula provided variations of the Lovelock couplings \cite{Kastor:2006vw} and shall not be considered here.  Since a black hole behaves as a thermodynamic system, the associated quantities should obey the first law which due to background quintessence modifies \cite{Chen:2008ra}: 
\begin{equation}\label{FL}
dM= T dS+ \Phi_q dq
\end{equation}
and for the constant $q$, it is same as $dM= T dS$.    Hence the entropy is given by
\begin{equation}
S=\int T^{-1}dM=\int T^{-1}\frac{dM}{dr}dr.
\end{equation}
Inserting Eq.~(\ref{T}) and the derivative of Eq.~(\ref{M}) into the above equation the entropy becomes 
\begin{equation}
S=2\pi(D-2)\Sigma_{D-2}\sum\limits^m_{s=1}\frac{\tilde{\alpha}_s\kappa^{s-1}s}{(D-2s)r_h^{-(D-2s)}}.\label{S}
\end{equation}
It is interesting to know that the mathematical form of the  entropy is same as entropy of EGB black holes. The contribution of background quintessence matter is reflected in the term $r_h$ of the entropy.  Eq.~(\ref{S}) suggests that in general the Lovelock black hole surrounded by quintessence matter does not obey the area law. However, in the special case $\kappa=0$ Eq.~(\ref{S}) reduces to
\begin{equation}
S=2\pi r^{D-2}_h\Sigma_{D-2}=2\pi A_{D-2}=\frac{A_{D-2}}{4G}, \label{SArea}
\end{equation}
where $A_{D-2}=r^{D-2}_h\Sigma_{D-2}$ is a horizon area of a $D$-dimensional black hole and in the last equation we have reinstated $8\pi G$. Hence we conclude that in spite of higher curvature terms the entropy of the black holes for $\kappa=0$ always obeys the area law \cite{Cai:2003kt}. For the limit $D=4$ it becomes a standard area law. The phase transition occurs in a  asymptotically AdS hairy Lovelock black holes \cite{Hennigar:2016ekz}.  Next we discuss the stability of black holes by computing heat capacity and the effect of surrounding quintessence matter. Thermodynamic stability of black holes is directly related to the sign of the heat capacity. The heat capacity is defined as 
\begin{equation}
C=\frac{d M}{d T}=\frac{d M}{dr_h}\frac{dr_h}{dT}.\label{Ct}
\end{equation}
Thus using Eqs.~(\ref{M}) and (\ref{T})
\begin{equation}
C=2\pi(D-2)\Sigma_{D-2}\frac{c_1c_2^2}{c_3},\label{C}
\end{equation}
where 
\begin{eqnarray}\nonumber
c_1&=&\sum\limits^m_{s=0}(D-2s-1)\frac{\tilde{\alpha}_s\kappa^{s}}{r^{-(D-2s-2)}_h}+\frac{q\omega_q(D-1)}{r_h^{(D-1)\omega_q+1}},~~
c_2=\sum\limits^m_{s=1}\frac{\tilde{\alpha}_s s\kappa^{s-1}}{r^{-(D-2s-1)}_h}. \\
c_3&=&c_2\partial_{r_h}c_1-c_1\partial_{r_h}c_2 
=\sum\limits^m_{s,p}(D-2p-1)(2s-2p-1)\frac{\tilde{\alpha}_s\tilde{\alpha}_p\kappa^{s+p-1}s}{r^{-2(D-s-p-2)}_h}\\
&-&\sum\limits^m_{s=1}[(D-1)\omega_q+1+(D-2s-1)]\frac{q\omega_q(D-1)}{r_h^{(D-1)\omega_q+1}}\frac{\tilde{\alpha}_ss\kappa^{s-1}}{r^{-(D-2s-2)}_h},\\\nonumber
\end{eqnarray}
One can see that the heat capacity depends on the Lovelock coefficients $\tilde{\alpha}_s$, the quintessence parameters $\omega_q$ and $q$ and the spacetime dimension $D$. It is known that the sign of the heat capacity is a criterion determining whether a thermodynamic system is stable or not. In other words, the positivity of heat capacity ensures that a thermodynamic system is stable while its negativity implies that a thermodynamic system is unstable. Thus, as black holes are considered as a thermodynamic system, for $C>0$ black holes are thermodynamically stable whereas for $C<0$ black holes are thermodynamically unstable. In Eq.~(\ref{C}) the limits $q\rightarrow0$, $\tilde{\alpha}_s\rightarrow0$ $(s\neq1)$ and $\tilde{\alpha}_1\rightarrow1$ with $\kappa = 1$ lead to the heat capacity for the Schwarzschild-Tangherlini black hole \cite{Ghosh:2014pga} 
\begin{equation}
C=-2\pi(D-2)\Sigma_{D-2}r^{D-2}.\label{STheatcapacity}
\end{equation}
The negative sign in Eq.~(\ref{STheatcapacity}) implies that the Schwarzschild-Tangherlini black hole is thermodynamically unstable \cite{Ghosh:2014pga}.
\begin{figure}[H]
	\begin{center}
		\begin{tabular}{cc}
			\includegraphics[scale=1.0]{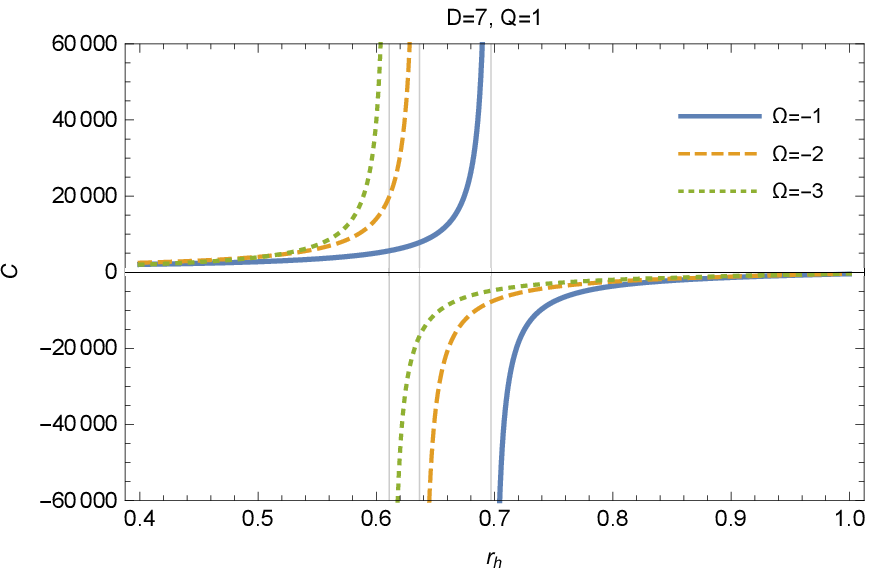} 
			&			\includegraphics[scale=1.0]{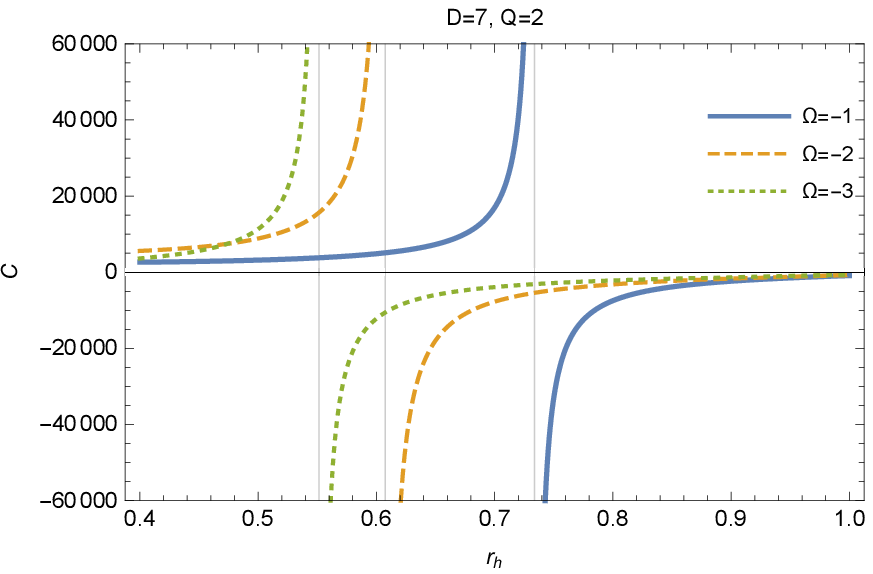}
		\end{tabular} 
	\end{center}
	\caption{The plots show how the heat capacity behaves for different values of $\omega_q$. We have chosen the parameters as $\kappa=1$, $D=7$, $\tilde{\alpha}_0=-1$ and $\tilde{\alpha}_1=\tilde{\alpha}_2=\tilde{\alpha}_3=1$. Here we used the parameters for $q$ and $\omega_q$, $Q\equiv -(D-1)q\omega_q$ and $\Omega\equiv(D-1)\omega_q$, respectively. The phase transition point $r_h=r_c$, where the heat capacity diverges increases as the $|\omega_q|$ increases.}\label{figComega}
\end{figure}
Eq.~(\ref{C}) suggests that the heat capacity depends on quintessence, and also spacetime dimensions. In the limit $q\rightarrow 0$ the heat capacity returns to the vacuum case \cite{Cai:2003kt}. In addition, in the limits $q\rightarrow0$, $\tilde{\alpha}_s\rightarrow0$ $(s\neq1)$ and $\tilde{\alpha}_1\rightarrow1$ with $\kappa = 1$ it goes to the general relativity case \cite{Ghosh:2014pga}. In order to bring out the effect of surrounding  quintessence on heat capacity we plot them (\textit{cf.} Figs.~\ref{figComega}, \ref{figCq} and \ref{figD}). The figures show that heat capacity diverges at the critical radius $r_c$ (\textit{cf.} Figs.~\ref{figComega}, \ref{figCq} and \ref{figD}). Further we know that heat capacity changes its sign around $r_h=r_c$. 
Thus the black hole is thermodynamically stable with positive heat capacity for $r_h<r_c$ and unstable, otherwise. We find the region where the Lovelock black holes is thermodynamically stable. In particular one finds that heat capacity of the Lovelock black hole can become positive for $r_h<r_c$ in all dimensions $D\geq7$ allowing the Lovelock black holes to achieve thermodynamic stability.


\section{Conclusion}
Lovelock gravity is one of  the most general  gravity theories in which the field equations are still second order and it is the natural generalization of  general relativity to higher dimensional spacetime. The basic idea is to supplement the Einstein-Hilbert action with the dimensionally continued Euler densities.  The Lovelock gravity has several additional interesting properties when compared with general relativity which triggered significant attention especially in finding black hole solutions in these theories. We have obtained an exact spherically symmetric black hole solution surrounded by quintessence matter in general Lovelock gravity thereby generalizing the vacuum solution for these theories. The current evidence for an accelerating early universe can be accommodated theoretically via a reintroduction of Einstein’s (positive) cosmological constant, which is equivalent to the introduction of quintessence matter with equation of state $p=- \rho$. More generally, and for general spacetime dimension D,  quintessence with equation of state $p=\omega_q \rho_q$ also yields a flat accelerating universe provided that $-1 \leq \omega_q \leq - 1 /(D-1)$. Despite complications of the geometry and horizon  of the black hole, we have found exact expressions for the thermodynamic quantities like the black hole mass, the Hawking temperature, entropy and heat capacity in terms of the horizon. In particular, we demonstrate that these thermodynamical quantities are corrected owing to quintessence term $q/r^{(D-1)(\omega_q+1)}$ in the solution Eq.~(\ref{PF}).   Explicit calculation of entropy shows that, in  general, the area law does not hold for the Lovelock black hole in Eq.~(\ref{PF}). One can understand the large scale structure of interactions containing quantum mechanical properties through thermodynamic quantities of black holes.  We confirmed that the entropy does not depend on the surrounding quintessence matter as found in \cite{Lee:2015xlp}.  In general, the phase transition occurs in an AdS black holes. However, with our parameter choice $\kappa=1$, $\tilde{\alpha}_0=-1$ and $\tilde{\alpha}_1=\tilde{\alpha}_2=\tilde{\alpha}_3=1$, we showed that the heat capacity has the phase transition point $r_c$  in various dimensions, where the heat capacity diverges. For a horizon radius below $r_c$ the heat capacity is positive, which means that the black hole is thermodynamically stable ,and beyond $r_c$ a thermodynamic unstable region starts, where the heat capacity is negative. Also we found that the phase transition point becomes larger as $|\omega_q|$ increases. This implies that in the equation of state  $p_q=\omega_q \rho_q$ when the magnitude of the pressure gets closer to that of the energy density, i.e. $\omega_q\rightarrow-1$, larger horizon radius range of stability is allowed, i.e. the black hole is more likely to be stable. On the other hand, we found that whether $r_c$ increases or decreases as $q$ increases also depends on $\omega_q$.  But in the particular case, $\kappa=0$, the area law is restored. In addition, thermodynamically stable black holes always appear with a positive heat capacity in all the dimensions. These thermodynamical properties are different from those in the general relativity. However, they become qualitatively similar to the Gauss-Bonnet black holes. Our result generalized previous approaches to a more general case, and in the limit $q\rightarrow 0$ this goes to the vacuum case.

\acknowledgements
S.G.G. would like to thank SERB-DST Research Project Grant No. SB/S2/HEP-008/2014 and DST INDO-SA bilateral project DST/INT/South Africa/P-06/2016 and also to IUCAA, Pune for the hospitality while this work was being done.

\end{document}